\documentclass[amsmath,amssymb,aps,prb,twocolumn]{revtex4-1}
\usepackage[dvips]{graphicx}
\usepackage{subfigure}
\usepackage{dcolumn}
\usepackage{bm}
\usepackage{hyperref}
\usepackage{array}
\usepackage{upgreek}
\usepackage[usenames]{color}
\usepackage[utf8]{inputenc}
\usepackage[T1]{fontenc}
\usepackage{mathptmx}
\usepackage{cancel}

\newcolumntype{P}[1]{>{\centering\arraybackslash}p{#1}}
\begin{document}
\preprint{AIP/123-QED}
\title{Interfacial Dzyaloshinskii-Moriya interaction in nonmagnetic/noncollinear-antiferromagnetic bilayers}
\author{Yuta Yamane,$^{1,2}$ Yasufumi Araki,$^3$ and Shunsuke Fukami$^{2,4,5,6,7,8}$}


\affiliation{$^1$Frontier Research Institute for Interdisciplinary Sciences, Tohoku University, Sendai 980-8578, Japan}
\affiliation{$^2$Research Institute of Electrical Communication, Tohoku University, Sendai 980-8577, Japan}
\affiliation{$^3$Advanced Science Research Center, Japan Atomic Energy Agency, Tokai 319-1195, Japan}
\affiliation{$^4$Advanced Institute for Materials Research, Tohoku University, Sendai 980-8577, Japan}
\affiliation{$^5$Graduate School of Engineering, Tohoku University, Sendai 980-0845, Japan}
\affiliation{$^6$Center for Science and Innovation in Spintronics, Tohoku University, Sendai 980-8577, Japan}
\affiliation{$^7$Center for Innovative Integrated Electronic Systems, Tohoku University, Sendai 980-0845, Japan}
\affiliation{$^8$Inamori Research Institute for Science, Kyoto 600-8411, Japan}

\date{\today}

\begin{abstract}
We study Dzyaloshinskii-Moriya interaction (DMI) appearing at the interface of a nonmagnetic/noncollinear-antiferromagnetic bilayer.
DMI is an antisymmetric exchange interaction between neighboring magnetic spins, arising in the absence of inversion center between the spins and the explicit expression of which being dictated by system symmetry.
We formulate the interfacial DMI for different crystalline orientations of the noncollinear antiferromagnet with stacked-Kagome lattice structure.
From this formulation, we show that, when the Kagome planes are perpendicular to the sample film plane,
the DMI serves as a uniaxial magnetic anisotropy for the antiferromagnetic order parameter.
Our findings reveal a novel physical manifestation of a DMI, shedding a new light on microscopic mechanisms of the magnetic anisotropy in noncollinear antiferromagnets.
\end{abstract}

\maketitle

\section{Introduction}
Dzyaloshinskii-Moriya interaction (DMI), an antisymmetric spin-spin exchange interaction resulting from spatial inversion breaking, plays fundamental roles in a variety of phenomena.
While originally proposed to understand the weak ferromagnetism exhibited by some antiferromagnets (AFMs) \cite{dmi1, dmi2}, it has recently been discussed that the DMI is also an essential ingredient in multiferroics \cite{mf1} and the magnon Hall effect \cite{mhe1}. 
The DMI has been gaining particularly renewed interest in the field of spintronics as a source of chiral spin textures such as magnetic skyrmions \cite{dmi_skyrm1, dmi_skyrm2, dmi_skyrm3, dmi_skyrm4} and chiral domain walls \cite{dmi_dw1, dmi_dw2, dmi_dw3}.
The effects of DMI are widely studied in nonmagnetic/magnetic bilayers \cite{idmi1, idmi2, idmi3, idmi4, idmi5}, where the DMI arises due to the breaking of inversion symmetry at the interface.
Such an interfacial DMI (i-DMI) can be tuned through interface engineering \cite{idmi-tune1}, strain \cite{idmi-tune2}, electric currents \cite{idmi-tune3, idmi-tune4} and electric fields \cite{idmi-tune5, idmi-tune6, idmi-tune7, idmi-tune8, idmi-tune9}.

Recently, the i-DMI arising in bilayers of a noncollinear AFM thin film and a nonmagnetic one has come to attention \cite{dmi-mn3sn-1, dmi-mn3sn-2, dmi-mn3sn-3, dmi-mn3sn-4, dmi-mn3sn-5}.
Here, noncollinear AFMs refer to frustrated Kagome AFMs, such as D0$_{19}$-Mn$_3$Sn \cite{mn3sn_1, mn3sn_2} and L1$_2$-Mn$_3$Ir \cite{mn3ir_1}, that show a triangular magnetic order.
In the emerging research field of antiferromagnetic spintronics \cite{af1, af2, af3}, noncollinear AFMs are expected to play pivotal roles;
Despite their vanishingly small net magnetization, they can exhibit anomalous Hall \cite{mn3sn_ahe1, mn3sn_ahe2, mn3sn_ahe3, mn3sn_ahe4}, magneto-optical \cite{mn3sn_moke1, mn3sn_moke2}, and anomalous Nernst \cite{mn3sn_ane1} effects whose magnitudes are comparable to those in ferromagnets.
They also host exotic spin transport phenomena \cite{mn3sn_tp1, mn3sn_tp2, mn3sn_tp3, mn3sn_tp4}.
Moreover, their AFM order can be electrically manipulated in field-assisted switching setups \cite{mn3sn_switch1, mn3sn_switch2, mn3sn_switch3, mn3sn_switch4, mn3sn_switch5, mn3sn_switch6, mn3sn_switch7, mn3sn_switch8} as well as through a novel type of spin excitation, namely the field-free continuous chiral spin rotation \cite{mn3sn_rot1, mn3sn_rot2, mn3sn_rot3}.
Recent successes in epitaxial growth of Mn$_3$Sn \cite{mn3sn_epi1, mn3sn_epi2, mn3sn_epi3} have propelled studies on noncollinear AFM thin films that are most often attached to some adjacent nonmagnetic films.
It has been hence becoming increasingly vital to understand the magnetization dynamics subject to the i-DMI at such interfaces.
The M-plane configuration, where the Kagome planes are perpendicular to the film plane, is of particular importance because it is a configuration suitable for studying electrical detection and manipulation of the magnetic state in a noncollinear AFM \cite{mn3sn_ahe1, mn3sn_ahe2, mn3sn_ahe3, mn3sn_ahe4, mn3sn_rot3, yamane2019}.
There lacks, however, a theoretical framework to quantitatively and systematically study physical implications of the i-DMI in such systems.
Since the interfacial structure in the M-plane configuration breaks the sublattice symmetry as well as the inversion symmetry, the framework for the i-DMI in ferromagnetic systems cannot be applied directly.

In this article, we present our theoretical formulation of the i-DMI for a nonmagnetic/noncollinear-AFM bilayer in the M-plane configuration.
We find that the i-DMI provides a sublattice-asymmetric renormalization to the bulk DMI, the latter being present due to the locally-broken inversion symmetry of the hexagonal Kagome lattice structure.
As a consequence of this i-DMI, a uniaxial anisotropy for the AFM order parameter with respect to the film normal direction is predicted to emerge.
Our analytical predictions are supported by numerical simulations of field- and current-induced dynamics of the AFM order.
These results are compared to those in the case of the C-plane configuration, where the Kagome planes lie parallel to the film plane \cite{dmi-mn3sn-2, dmi-mn3sn-5}, confirming that the analytical expression and physical implications of the i-DMI depend crucially on the crystalline orientation.

\section{Starting Atomistic Model}
\begin{figure}[t]
\centering
\includegraphics[width=8.5cm, bb=0 0 1194 643]{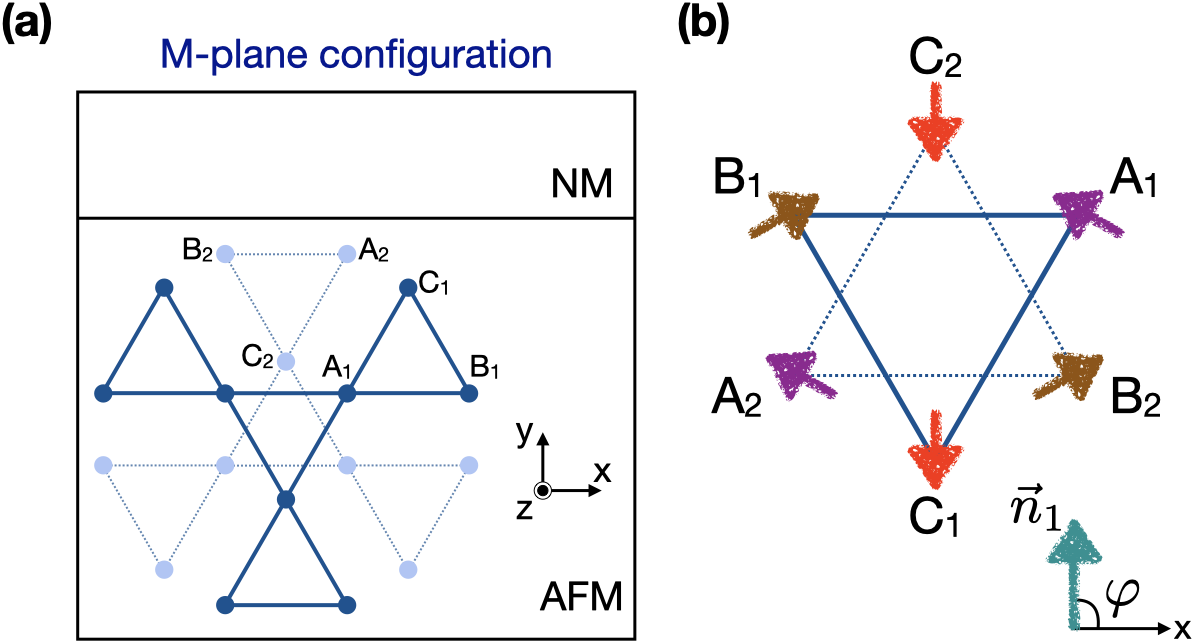}
\caption{
(a) Schematic of our starting atomistic model;
a bilayer of a nonmagnet (NM) and a stacked-Kagome antiferromagnet (AFM) in the M-plane configuration.
(b) A crystalline unit cell of the AFM, where the arrows represent the magnetic moments.
$ \vec n_1 $ is one of the two N\'eel vectors defined in the coarse-grained, three magnetic-sublattice model, and $ \varphi $ is the angle of $ \vec n_1 $ in the $ xy $ plane.
The magnetic-moment structure drawn here corresponds to $ \varphi = \frac{ \pi }{ 2 } $.
}
\label{fig01}
\end{figure}

We consider a bilayer of a nonmagnet and a noncollinear AFM, in the latter of which the magnetic atoms form a stacked-Kagome lattice structure as in D0$_{19}$-Mn$_3$Sn \cite{mn3sn_1, mn3sn_2} and L1$_2$-Mn$_3$Ir \cite{mn3ir_1}.
We primarily focus in this article on the so-called M-plane configuration, where the Kagome planes are oriented perpendicular to the film plane [Fig.~1~(a)].
The coordinate system is set so that the $ z $ axis is normal to the Kagome planes.
We take a unit cell of the AFM as it contains six magnetic atoms from two adjacent Kagome layers [Fig.~1~(b)].
We label the three atoms on the layer 1 (represented by the solid lines in Fig.~1) as $ A_1 $, $ B_1 $ and $ C_1 $, while the other three on the layer 2 (represented by the dotted lines) as $ A_2 $, $ B_2 $ and $ C_2 $.
The entire AFM thin film is made of the repetitive 1212... stacking of the Kagome layers.
In the unit cell chosen as in Fig.~1~(b), the lattice vectors $ \vec d_{ ij } $, pointing from the atom $ j $ to $ i $ within the same Kagome layer, are given by $ \vec d_{ B_1 A_1 } = - \vec d_{ B_2 A_2 } = - a \vec e_x $, $ \vec d_{ C_1B_1 } = - \vec d_{ C_2 B_2 } = a \vec e_{ - \frac{ \pi }{ 3 } } $, and $ \vec d_{ A_1 C_1 } = - \vec d_{ A_2 C_2 } = a \vec e_{ \frac{ \pi }{ 3 } } $, where $ a $ is the intralayer lattice constant, $ \vec e_{ \mu = x, y, z } $ is the unit vector in the $ \mu $ axis, and $ \vec e_\phi = \cos\phi \vec e_x + \sin\phi \vec e_y $.
The vectors connecting the interlayer atoms in the unit cell, on the other hand, are given by $ \vec d_{ B_2 A_1 } = - \vec d_{ B_1 A_2 } = - \frac{ a }{ \sqrt{ 3 } } \vec e_y - \sqrt{ \frac{ 2 }{ 3 } } c \vec e_z $, $ \vec d_{ C_2 B_1 } = - \vec d_{ C_1 B_2 } = \frac{ a }{ \sqrt{ 3 } } \vec e_{ \frac{ \pi }{ 6 } } - \sqrt{ \frac{ 2 }{ 3 } } c \vec e_z $, and $ \vec d_{ A_2 C_1 } = - \vec d_{ A_1 C_2 } = \frac{ a }{ \sqrt{ 3 } } \vec e_{ \frac{ 5 \pi }{ 6 } } - \sqrt{ \frac{ 2 }{ 3 } } c \vec e_z $, where $ c $ is the interlayer spacing.
The nonmagnetic film is not explicitly incorporated in the model;
We introduce its effect as the lowering of the symmetry at the interface, compared to the symmetry in the bulk AFM.

The DMI energy $ U_{\rm dmi} $ is introduced by
\begin{equation}
	U_{\rm dmi} = \sum_{ ( i, j ) } \left( \vec D_{ ij }^{\rm bulk} + \vec D_{ ij }^{\rm int} \right)
	                       \cdot \vec \mu_i \times \vec \mu_j ,
\label{dmi}
\end{equation}
where $i$ and $j$ denote the lattice sites, $ \vec \mu_i $ is the classical magnetic moment located at $ i $, and the summation is taken over the pairs of nearest-neighbor intra- and intersublattice sites $ ( i, j ) $.
The bulk DMI vector $ \vec D_{ ij }^{\rm bulk} $ originates from the locally-broken inversion symmetry of the intrinsic crystalline structure; i.e., while the stacked-Kagome lattice as a whole has an inversion center, a given pair of two sites does not necessarily has one in between them.
The i-DMI vector $ \vec D_{ ij }^{\rm int} $, on the other hand, is induced by the globally-broken inversion symmetry due to the existence of the interface.
The magnitude $ | \vec D_{ ij }^{\rm int} | $ should take an appreciable value only for pairs $(i,j)$ sufficiently close to the interface, while $ | \vec D_{ ij }^{\rm bulk} | $ should be almost insensitive to the distance from the interface.  
We don't need to know, however, the exact spatial dependence of the DMI vectors for the discussion below.
The directions of the DMI vectors are given by
\begin{equation}
	\vec D_{ ij }^{\rm bulk} \propto \vec d_{ ij } \times \vec c_{ ij } , \qquad
	\vec D_{ ij }^{\rm int} \propto \vec d_{ ij } \times \vec t ,
\label{d_vec}
\end{equation}
with $ \vec c_{ ij } $ the unit vector pointing from the inversion center of the unit cell toward the link $ij$,
and $\vec t $ the unit vector in the film normal direction \cite{kagome-dmi}.
In the M-plane configuration considered here, we take $ \vec t = \vec e_y $.
We discuss in Section V the configuration where the Kagome planes lie in the sample film (C-plane configuration), where we take $ \vec t = \vec e_z $.

\section{Analysis with continuum approximation}
We here take on a continuum approximation,
by performing a coarse graining onto the atomistic model introduced in the last section.
In doing so, we adopt a three magnetic-sublattice model, where we ignore the layer degree of freedom, unifying the six sublattices $(A_1,B_1,C_1,A_2,B_2,C_2)$ into the three $(A,B,C)$.
The magnetic moments from the atoms $ A_1 $ and $ A_2 $ are collectively represented by the classical, unit-vector continuous field $ \vec m_A ( \vec r, t ) $, and similarly for $ \vec m_B ( \vec r, t ) $ and $ \vec m_C ( \vec r, t ) $.

In the rest of this and next sections, we assume the spatial homogeneity in the mean field, $ \vec m_\zeta ( \vec r, t ) = \vec m_\zeta ( t ) $, for each sublattice $ \zeta = A, B, C $.
The assumption of the homogeneity along the film-normal ($y$) direction is justified when the film thickness is comparable to or shorter than the exchange length of the AFM film.
This condition is satisfied with an epitaxial Mn$_3$Sn thin film with 10 - 100 nm of the thickness \cite{mn3sn_epi3}.
Under the assumption of the spatial homogeneity, the lattice translational symmetry is therefore fully retained along the $z$ and $x$ axes.
Spatially inhomogeneous magnetizations varying in the film plane ($zx$ plane) are touched upon in Section V.

We now derive the expression of the DMI energy density in the coarse-grained, three magnetic-sublattice model.
The general expression of the DMI energy density, including both the bulk and interfacial DMIs, is written as
\begin{equation}
	u_{\rm dmi} = \vec D_{AB} \cdot \vec m_A \times \vec m_B 
	                       + \vec D_{BC} \cdot \vec m_B \times \vec m_C
	                       + \vec D_{CA} \cdot \vec m_C \times \vec m_A ,
\label{dmi_general}
\end{equation}
where $ \vec D_{ \zeta \eta} ( \zeta, \eta = A, B, C ) $ are the DMI vectors in the continuum model.
Note that now $ \vec D_{ \zeta \eta} $ do not depend on space as a result of the coarse graining.
We are considering an AFM thin film where its thickness is sufficiently small that a good portion of the AFM is subject to the appreciably large i-DMI and, after averaging the atomistic i-DMI over the film-normal ($y$) direction for each sublattice-pair, the contribution of the i-DMI to $ \vec D_{ \zeta \eta } $ is still discernible along with the bulk counterpart.

\begin{figure}[b]
\centering
\includegraphics[width=8cm, bb=0 0 867 580]{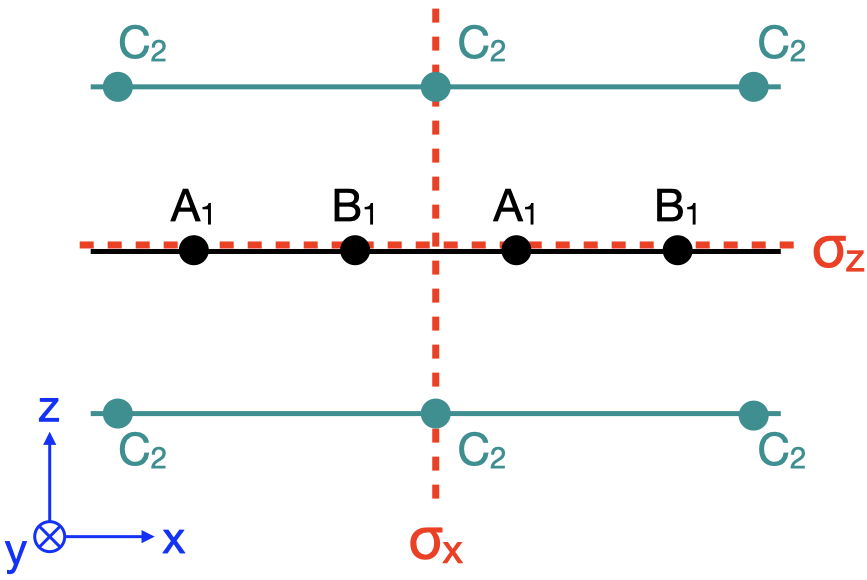}
\caption{
Symmetry operations in the stacked-Kagome lattice are schematically shown, where mirror planes for $ \sigma_z $ and $ \sigma_x $ are indicated by the dotted lines.
For clarity, we only show a chain of A$_1$ and B$_1$ sites in the middle plane and chains of C$_2$ sites in the bottom and upper planes.
}
\label{fig02}
\end{figure}

The relation among the three $ \vec D_{ \zeta \eta}$'s are determined by the symmetries of the original atomistic model.
By taking into account the interfacial structure, we consider two mirror reflections (Fig.~2).
The first is the reflection by an arbitrary Kagome plane ($xy$ plane), which we denote as $\sigma_z$.
In the coarse-grained model, $ \sigma_z $ corresponds to reversing the three sublattice magnetizations as $ ( m_\zeta^x, m_\zeta^y, m_\zeta^z ) \rightarrow ( - m_\zeta^x, - m_\zeta^y, m_\zeta^z ) $ $ \forall \zeta $ while it does not change the sublattice indices.
Under this operation, the DMI energy density transforms as $ u_{\rm dmi} \rightarrow \sigma_z [ u_{\rm dmi} ] \equiv \sum_{ ( \zeta \eta ) } [ - D^x_{ \zeta \eta } ( \vec m_\zeta \times \vec m_\eta )_x - D^y_{ \zeta \eta } ( \vec m_\zeta \times \vec m_\eta )_y + D^z_{ \zeta \eta } ( \vec m_\zeta \times \vec m_\eta )_z ] $.
Since the underlying lattice structure (in the presence of the interface) is symmetric with respect to $ \sigma_z $, $ u_{\rm dmi} $ must be invariant under the $ \sigma_z $ operation;
it is clear from the comparison of $ u_{\rm dmi} $ and $ \sigma_z [ u_{\rm dmi} ] $ that 
only the $ z $ components of $ \vec D_{ \zeta \eta } $ are allowed to be nonzero.
The second mirror operation is the reflection by a $yz$ plane passing through either $ C_1 $ or $ C_2 $ atoms, which we denote as $\sigma_x$.
In the coarse-grained model, this operation yields $ ( m_\zeta^x, m_\zeta^y, m_\zeta^z ) \rightarrow ( m_{\zeta'}^x, - m_{\zeta'}^y, - m_{\zeta'}^z ) $, along with the sublattice mapping $(\zeta,\zeta') = (A,B),(B,A),(C,C) $.
The required invariance of $ u_{\rm dmi} $ under the $\sigma_x$ operation now demands $ D_{ BC }^z = D_{ CA }^z $, whereas $D_{AB}^z$ is left independent.
In the case of bulk AFM, i.e., when the AFM fills the whole space with no interface existing, the threefold rotational symmetry of the stacked-Kagome lattice would be further retained, and hence the DMI becomes highly symmetric, $D_{ BC }^z = D_{ CA }^z = D_{ AB }^z$.
From these symmetry considerations, we can split $u_{\rm dmi}$ into the bulk and interfacial contributions, $ u_{\rm dmi} = u_{\rm bdmi} + u_{\rm idmi} $, with
\begin{eqnarray}
	u_{\rm bdmi} &=& D_0 \vec e_z \cdot \left( \vec m_A \times \vec m_B 
	                                                                    + \vec m_B \times \vec m_C
	                                                                    + \vec m_C \times \vec m_A \right) ,
	                             \label{bdmi} \\
	u_{\rm idmi} &=& D \vec e_z \cdot
	                             \left( \vec m_B \times \vec m_C + \vec m_C \times \vec m_A \right) .
	                             \label{idmi}
\end{eqnarray}
Here $ D_0 $ and $ D $ are phenomenological bulk and interfacial DMI constants, respectively.
While the microscopic i-DMI in Eq.~(\ref{dmi}) may also contain a sublattice-symmetric part, we can always absorb it into $ u_{\rm bdmi} $.
Thus, only the $BC$ and $CA$ couplings are left in $ u_{\rm idmi} $,
representing a sublattice-asymmetric modulation to the bulk DMI.
In the C-plane configuration, on the other hand, the structure of the i-DMI is completely different (see Section V).

To examine physical effects of the i-DMI, we consider a noncollinear AFM modeled by the following total energy density,
\begin{eqnarray}
	u &=& J \sum_{ ( \zeta \eta ) } \vec m_\zeta \cdot \vec m_\eta
                   + u_{\rm bdmi} + u_{\rm idmi}
                   \nonumber \\ &&
 	           - K \sum_{ \zeta = A, B, C } \left( \vec e_{ K_\zeta } \cdot \vec m_\zeta \right)^2
	           - \mu_0 M_S^{\rm sub} \vec h \cdot \sum_{ \zeta = A, B, C } \vec m_\zeta .
\label{u}
\end{eqnarray}
The first term represents the AFM exchange coupling with the coupling constant $J(>0)$, where $ ( \zeta \eta ) $ indicates summation over the pairs $(A,B)$, $(B,C)$ and $(C,A)$.
The second and third terms are the bulk and interfacial DMIs, as introduced in Eqs.~(\ref{bdmi}) and (\ref{idmi}), respectively.
The fourth term with the coefficient $K(>0)$ represents the local uniaxial anisotropy originating from the hexagonal crystalline structure, with the anisotropy axes $ \vec e_{ K_A } = \vec e_{ \frac{ 2 \pi }{ 3 } } $, $ \vec e_{ K_B } = \vec e_{ \frac{ 4 \pi }{ 3 } } $, and $ \vec e_{ K_C } = \vec e_x $ for the three magnetic sublattices.
The last term comes from the Zeeman coupling to an external magnetic field $\vec h$, where $ M_S^{\rm sub} $ is the saturation magnetization of each sublattice.

Let us introduce the two sublattice-asymmetric order parameters \cite{mn3sn_rot1, yamane2019}, $ \vec n_1 = \frac{ \vec m_A + \vec m_B - 2 \vec m_C }{ 3 \sqrt{ 2 } } $ and $ \vec n_2 = \frac{ - \vec m_A + \vec m_B }{ \sqrt{ 6 } } $, and the net magnetization $ \vec m = \frac{ \vec m_A + \vec m_B + \vec m_C }{ 3 } $.
As a direct consequence of these definitions, we find the relations, $ \vec n_1^2 + \vec n_2^2 + \vec m^2 = 1 $, $ \vec m \cdot \vec n_1 = \frac{ \vec n_1^2 - \vec n_2^2 }{ 2 \sqrt{ 2 } } $, and $ \vec m \cdot \vec n_2 = - \frac{ \vec n_1 \cdot \vec n_2 }{ \sqrt{ 2 } } $.
We now assume $ J \gg | D_0 | \gg K \gg | D | $, a condition that guarantees that the sublattice magnetizations make an as nearly-perfect triangular configuration in the $ xy $-plane \cite{mn3sn_2,yamane2019} as $ | \vec m | \ll 1 $ and $ | m_\zeta^z | \ll 1 $ $ \forall \zeta $;
the AFM exchange coupling ensures the three sublattice-magnetizations to make the relative angle of $\sim120^\circ$ with each other, and then the Kagome ($xy$) plane is chosen as the easy plane by the bulk DMI and the local anisotropy.
(The bulk DMI also dictates the chirality of the $120^\circ$ rotation of the magnetizations, as elaborated shortly.)
The i-DMI is assumed to be perturbatively small compared to the other energy terms, so that the AFM remains well described by the triangular order in the Kagome plane.
The above condition leads straightforwardly to
$ n_1^z \simeq n_2^z \simeq 0 $, $ | \vec n_1 | \simeq | \vec n_2 | \simeq \frac{ 1 }{ \sqrt{ 2 } } $, and $ \vec n_1 \cdot \vec n_2 \simeq 0 $.
It is then plausible to describe $ \vec n_1 $ and $ \vec n_2 $ by a single parameter \cite{nafm1, nafm2} as $ \vec n_1 = \frac{ 1 }{ \sqrt{ 2 } } ( \vec e_x \cos\varphi + \vec e_y \sin\varphi ) $ and $ \vec n_2 = R_\pm \vec n_1 \equiv \frac{ 1 }{ \sqrt{ 2 } } \left[ \vec e_x \cos( \varphi \pm \frac{ \pi }{ 2 } ) + \vec e_y \sin ( \varphi \pm \frac{ \pi }{ 2 } ) \right] $, where $\varphi$ can be interpreted as corresponding to the direction of the magnetic octupole moment \cite{mn3sn_ahe4}.
The bulk DMI energy favors $ R_- $ ($ R_+ $) when $ D_0 $ is positive (negative), while the local anisotropy energy can be minimized only with $ R_+ $.
The inverse triangular structure observed in, e.g., Mn$_3$Sn, is realized with $ R_- $, while the all-in/all-out triangular structure seen in, e.g., Mn$_3$Ir, is realized with $ R_+ $.
Assuming now $ D_0 > 0 $, we choose $ R_- $, based on the condition $ D_0 \gg K $ placed before.
As a consequence of the conflict between the two energies, the magnetic structure exhibits a slight deviation from a perfect triangular configuration, which results in a nonzero net magnetization $ \vec m $ appearing in the Kagome plane in the ground states \cite{mn3sn_2,yamane2019}.
This is the case for Mn$_3$X (X = Sn, Ge, Ga) \cite{mn3sn_1, mn3sn_2, mn3ga_1, mn3ga_2}.

Now we consider expressing the i-DMI energy density $ u_{\rm idmi} $ in terms only of $ \varphi $, under the condition assumed above, namely $ J \gg D_0 \gg K \gg | D | $ that assures $ | \vec m | \ll 1 $ and $ n_1^z \simeq n_2^z \simeq 0 $.
In doing so, we first rewrite Eq.~(\ref{idmi}) in terms of $( \vec n_1, \vec n_2, \vec m ) $ simply using their definitions given above.
Then, $ \vec n_1 $ and $ \vec n_2 $ are expressed in terms of $ \varphi $ as discussed above.
As to $ \vec m $, it has been shown that $ \vec m $ can be given as a function of $ \vec n_1 $ (or equivalently $ \vec n_2 $ using the relation $ \vec n_2 = R_- \vec n_1 $) as long as the condition $ | \vec m | \ll 1 $ is satisfied \cite{mn3sn_rot1, yamane2019}.
In equilibrium, it is given at the first order of the small quantity $ J^{-1} $ as \cite{yamane2019}
$ \vec m = \frac{ 1 }{ 3 J } \left[ \mu_0 M_S^{\rm sub} \vec h + K \left( - \vec e_x \cos\varphi + \vec e_y \sin\varphi \right) \right] $.
The first term describes the disturbance to the triangular structure due to the external filed.
(Here, of course, the field is assumed to be sufficiently small that the Zeeman energy is small compared to the exchange coupling energy.)
The second term is the intrinsic canting originating from the local anisotropy conflicting with the $ R_- $ chirality, as discussed above.
We ignored possible additional contributions to $ \vec m $ from the i-DMI, because such terms would lead to second or higher order terms of $D$ in $ u_{\rm idmi} $, and they are negligible in the present perturbative treatment of the i-DMI.
With those expressions of $( \vec n_1, \vec n_2, \vec m ) $ in terms of $ \varphi $ substituted into $ u_{\rm dmi} $, one is led to
\begin{equation}
	u_{\rm idmi} = \frac{ 2 K }{ \sqrt{ 3 } J } D \sin^2\varphi
	                       + \frac{ \mu_0 M_S^{\rm sub} }{ \sqrt{ 3 } J } D \left( h_x \cos\varphi + h_y \sin\varphi \right) ,
\label{u_idmi}
\end{equation}
where we have discarded the terms independent of $ \varphi $.

We find that the first term in Eq.~(\ref{u_idmi}) serves as a uniaxial anisotropy for the AFM order direction $ \varphi $, which is the central result of this work.
Even though an interfacial magnetic anisotropy is not microscopically present for each magnetic moment in our current model, the uniaxial anisotropy effectively arises from the i-DMI at the macroscopic scale.
The second term in Eq.~(\ref{u_idmi}) has a form of the Zeeman coupling between the external field $\vec h$ and the net magnetization $\vec m$, which serves as a renormalization to the built-in Zeeman coupling term in Eq.~(\ref{u}).
In what follows we examine cases with $ D < 0 $;
noting that $ \varphi $ is measured from the $ x $ axis, the negative $ D $ makes the $ y $ axis, the film-normal direction, the easy axis for $ \varphi $.
If $ D > 0 $, on the other hand, the easy axis would be along the $ x $ axis.

The nonzero $ \vec m $ in the ground states allows for magnetic-field control of $ \varphi $.
Let us consider the switching between $ \varphi = + \frac{\pi}{2} $ and $ - \frac{\pi}{2} $ by an external field applied along the $ y $ axis [Fig.~3~(a)].
{The expression of the total magnetic energy density $ u $ in terms of $\varphi$ can now be approximately given, at the first order of $ J^{-1} $, as
\begin{equation}
	u \simeq - K_{\rm idmi} \sin^2\varphi - \mu_0 M_{\rm idmi} h_y \sin\varphi ,
\label{u_tot}
\end{equation}
with
\begin{equation}
	K_{\rm idmi} = \frac{ 2 K }{ \sqrt{ 3 } J } | D | , \qquad
	M_{\rm idmi} = \frac{ M_S^{\rm sub} }{ J } \left( K +  \frac{ | D | }{ \sqrt{ 3 } } \right) .
\end{equation}
Here, $K_{\rm idmi}$ describes the effective uniaxial anisotropy, and $M_{\rm idmi}$ parametrizes the renormalized Zeeman coupling, both of which are consequences of the i-DMI.
Note that the bulk DMI in Eq.~(\ref{u}) only gives a constant energy under the condition $ \vec n_1 \cdot \vec n_2 \simeq 0$.
The critical field $ h_y^* $ for the switching of $ \varphi $ can then be obtained by applying the standard Stoner-Wohlfarth approach \cite{sw} to Eq.~(\ref{u_tot}), the result of which is
\begin{equation}
	| h_y^* | = \frac{ 2 K_{\rm idmi} }{ \mu_0 M_{\rm idmi } }
	              = \frac{ 4 | D | }{ \mu_0 M_S^{\rm sub} } \frac{ K }{ \sqrt{ 3 } K + | D | } .
\label{hc}
\end{equation}
The critical field is not simply linear with $ | D | $, because the i-DMI contributes not only to the anisotropy via $ K_{\rm idmi} $ but also to the effective renormalization of the Zeeman energy via $ M_{\rm idmi} $.

In addition to the uniaxial anisotropy of the i-DMI origin, the hexagonal crystalline structure of the AFM allows an intrinsic, sixfold anisotropy for $ \varphi $.
This sixfold anisotropy, however, appears at the order of $ J^{-2} $ in the perturbative expansion of the magnetic energy density \cite{balents}, and hence is not included in Eq.~(\ref{u_tot}).
In the bulk AFM, where $ D = 0 $ and thus the sixfold anisotropy is the leading-order anisotropy, the energy minima (maxima) correspond to $ \varphi = 0, \pi, \pm \frac{ \pi }{ 3 }, \pm \frac{ 2 \pi }{ 3 } $ $ \left( \varphi = \pm \frac{ \pi }{ 6 }, \pm \frac{ \pi }{ 2 } , \pm \frac{ 5 \pi }{ 6 } \right) $.
Once the i-DMI is introduced $ ( D \neq 0 ) $, the effective uniaxial anisotropy of $O(J^{-1})$ dominates over this sixfold one as we numerically verify in the next section.

\section{Dynamical Simulation}
\begin{figure}[t]
\centering
\includegraphics[width=8.5cm, bb=0 0 1561 977]{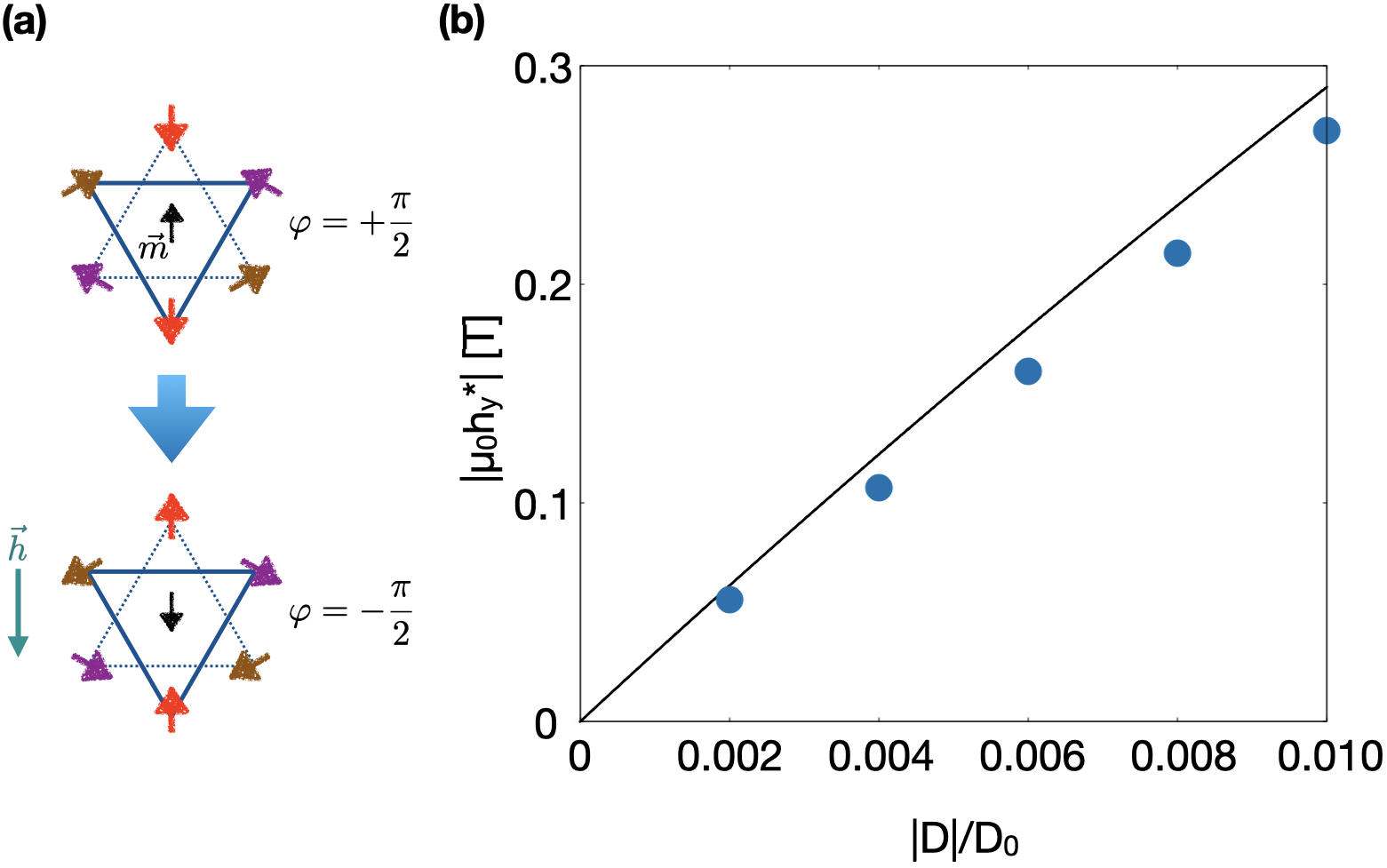}
\caption{
(a) Schematic of the switching from $ \varphi = + \frac{\pi}{2} $ to $ - \frac{\pi}{2} $, by the magnetic field $ \vec h $ applied in the $ - y $ direction.
$ \vec m $ is the net magnetization due to the slight deviation from a perfect triangular structure, which allows the field switching.
(b) The absolute value of the switching field $ | \mu_0 h_y^* | $ against the absolute value of the i-DMI constant $ | D | $ (normalized by the bulk DMI constant $ D_0 $) in the M-plane configuration.
The solid curve plots Eq.~(\ref{hc}), and the circle symbols are results of the LLG simulations.
See the main text for the parameter values employed.
}
\label{fig03}
\end{figure}

To check the theoretical predictions derived in the previous section, we now numerically simulate the field-induced switching of the AFM order.
The sublattice magnetizations are subject to the coupled Landau-Lifshitz-Gilbert (LLG) equations,
\begin{equation}
	\partial_t \vec m_\zeta = - \vec m_\zeta \times \left( \gamma \vec h_\zeta + \alpha \partial_t \vec m_\zeta \right) ,
                                                \quad (\zeta=A,B,C) ,
\label{llg}
\end{equation}
where the effective fields $ \vec h_\zeta $ are defined by $ \vec h_\zeta = - ( \mu_0 M_S^{\rm sub} )^{-1} \partial u / \partial \vec m_\zeta $, $\gamma $ is the gyromagnetic ratio, and $ \alpha $ is the Gilbert damping constant.
We set the parameter values so as them to be in a reasonable range for a representative noncollinear AFM Mn$_3$Sn \cite{mn3sn_epi3}, where the condition $ J \gg D_0 \gg K $ is satisfied;
$ \mu_0 M_S = 0.55$ T, $ J = 6.2 \times10^7 $ Jm$^{-3} $, $ D_0 = 6 \times 10^6 $ Jm$^{-3}$, $ K = 3.8 \times 10^5 $ Jm$^{-3}$, and $ \alpha = 0.001 $.
To examine the effects of the i-DMI,  $D$ is varied in the range $0 < |D| \leq 0.01 D_0$.
We start from the initial state with $ \varphi = + \frac{ \pi }{ 2 } $, corresponding to the net magnetization $ \vec m $ pointing in the $ + y $ axis.
We then apply a magnetic field $\vec h = -|h_y| \vec e_y$ along the $ - y $ direction.
By sweeping the field strength $|h_y|$ from zero, we detect the switching field $|h_y^*|$ where the AFM order switches to $ \varphi = - \frac{ \pi }{ 2 } $ ($ \vec m $ in the $ - y $ direction), as schematically depicted in Fig.~3~(a).

The numerically-obtained switching field $|h_y^*|$ is compared with the analytical relation Eq.~(\ref{hc}), as plotted against $ | D | $ in Fig.~3~(b).
The numerical results show a quantitatively good agreement with the analytical relation, especially in the small $ | D | $ regime where the perturbative treatment of the i-DMI is more appropriate.
From both the analytical and numerical results, we clearly see that the i-DMI gives rise to a uniaxial anisotropy for $ \varphi $, whose strength monotonically increases with $ | D | $.
We note that the switching field can reach as large as $ | \mu_0 h_y^*| \sim 270 $ mT, even when the strength of the i-DMI $|D|$ is only one percent that of the bulk DMI $D_0$.
Such a value of the switching field is comparable to those experimentally reported for Mn$_3$Sn thin films \cite{mn3sn_tp4, mn3sn_switch3, mn3sn_switch7, mn3sn-ani1, mn3sn_epi3}.
In the previous study, the experimentally-observed uniaxial anisotropy has been attributed to the strain-induced sublattice-asymmetry in the AFM exchange coupling \cite{mn3sn_switch3, mn3sn_switch7, mn3sn-ani1}. 
Our simulation suggests that the i-DMI may also give an appreciable contribution to the anisotropy, revealing a new route to manipulating magnetic anisotropy through i-DMI engineering.

In the case of $ D = 0 $, the perpendicular states $ \varphi = \pm \frac{ \pi }{ 2 } $ are not the ground states under the intrinsic sixfold anisotropy, and switching would be between the six ground states mentioned above.
We numerically find that, when $ D = 0 $,  the switching field $ | \mu_0 h_y^*| $ for a ``up-to-down'' switching from $ \varphi = + \frac{ \pi }{ 3 } $ to $ - \frac{ \pi }{ 3 } $ is about 8 mT.
The intrinsic sixfold anisotropy is thus negligible compared to the uniaxial one originating from the i-DMI.

\section{Discussions}
\begin{figure}[b]
\centering
\includegraphics[width=8cm, bb=0 0 889 483]{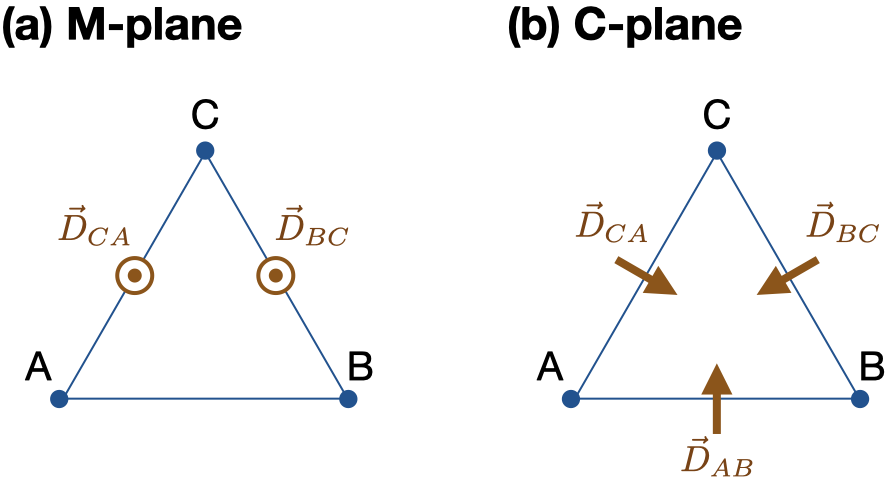}
\caption{
Relationships of the i-DMI vectors in the three-sublattice model, with (a) the M-plane [Eq.~(\ref{idmi})] and (b) the C-plane [Eq.~(\ref{idmi_c})] configurations.
Note that, in the three-sublattice model, the magnetizations are coarse-grained, continuous fields in space, i.e., there are no discrete $ (A,B,C)$ ``sites'' in the model. 
The triangles drawn here are just for an easy-to-grasp, conceptual visualization of the relationship of the i-DMI vectors among the three sublattices, not representing the actual model structure in the real space.
} 
  \label{fig04}
\end{figure}
\begin{figure}[t]
\centering
\includegraphics[width=8.5cm, bb=0 0 1547 977]{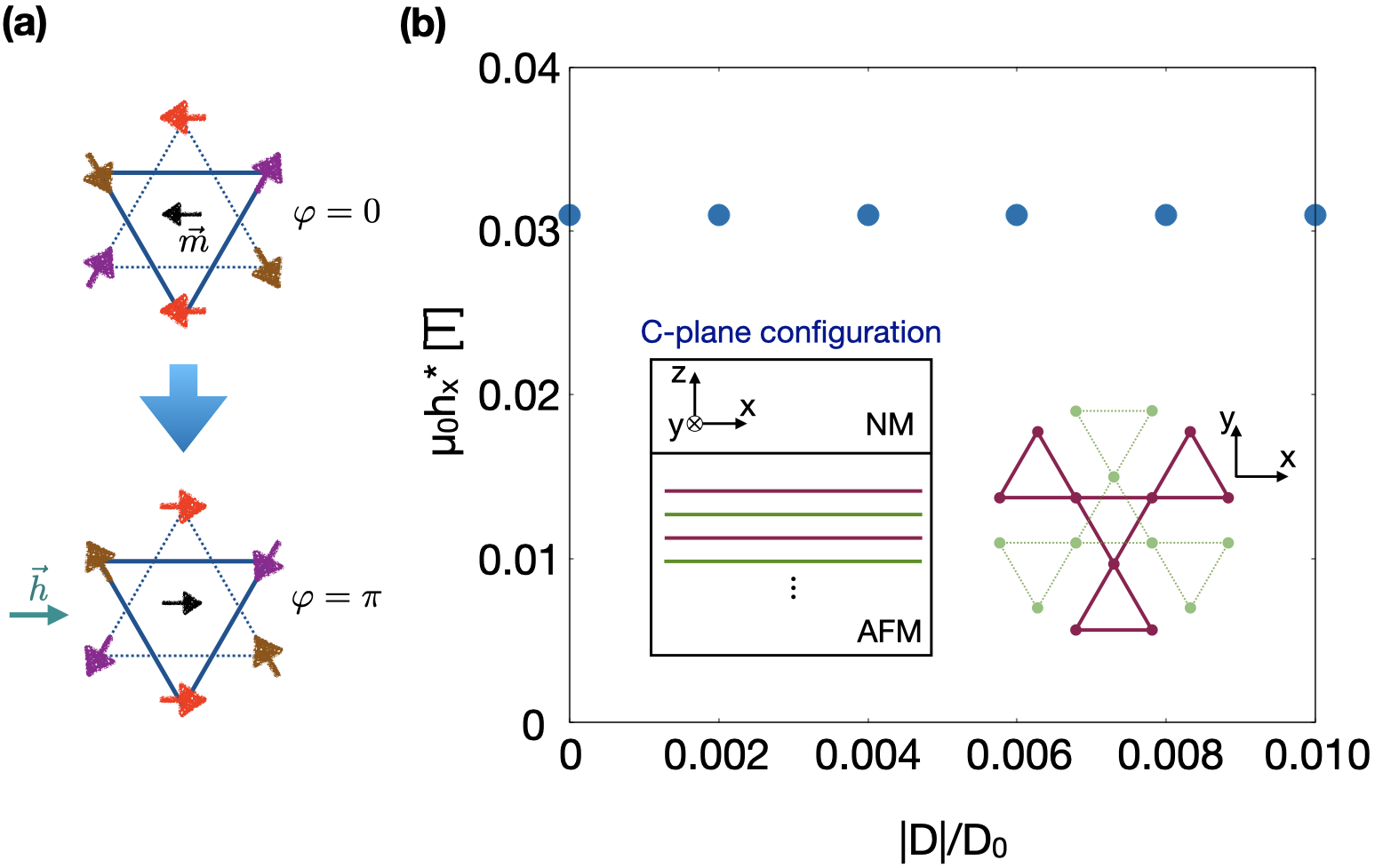}
\caption{
(a) Schematics depicting the switching from $ \varphi = 0 $ to $ \pi $ by the magnetic field $ \vec h $ applied in the $ + x $ direction.
  (b) The numerically-obtained switching field $ \mu_0 h_x^* $, plotted against the absolute value of the i-DMI constant $ | D | $ (normalized by the bulk DMI constant $ D_0 $) in the C-plane configuration.
  See the main text for the parameter values employed.
  In the inset, our atomistic starting model for the C-plane configuration is schematically shown.
                               }
  \label{fig05}
\end{figure}
In order to see that the crystalline orientation is crucial for physical manifestations of the i-DMI, we here briefly discuss the C-plane configuration, where we take $ \vec t = \vec e_z $.
The bulk part $ u_{\rm bdmi} $ of the DMI energy density is independent of the crystalline orientation with respect to the interface, so that it is given by Eq.~(\ref{bdmi}) in the C-plane configuration as well.
The relations among the i-DMI vectors, on the other hand, are determined by the particular symmetry of the C-plane configuration.
We start from the general expression of the DMI in Eq.~(\ref{dmi_general}), and consider the mirror operation $ \sigma_x $ (introduced before) and the rotational operation $ C_3 $: 
$ \vec m_\zeta \rightarrow R_3 \vec m_\zeta \equiv \left( m_\zeta^x \cos\frac{ \pi }{ 3 } - m_\zeta^y \sin\frac{ \pi }{ 3 } , m_\zeta^x \sin\frac{ \pi }{ 3 } + m_\zeta^y \cos\frac{ \pi }{ 3 } , m_\zeta^z \ \right) $ accompanied by the sublattice mapping $ ( \zeta , \zeta' ) = ( A, C ) , ( B, A ), ( C, B ) $.
The requirement that the i-DMI energy be invariant under $ C_3 $ leads to $ D_{ AB}^z = D_{ BC}^z = D_{ CA}^z $ and $ \vec D_{ AB }^\parallel = R_3 \vec D_{ CA }^\parallel = R_3^2 \vec D_{ BC }^\parallel $, where $ \vec D_{ \zeta \eta }^\parallel $ is the component of $ \vec D_{ \zeta \eta } $ in the Kagome $( xy )$ plane.
The invariance under the $ \sigma_x $ operation, on the other hand, requires $ D_{ AB }^x = 0 $, $ D_{ BC }^x = - D_{ CA }^x $, and $ D_{ BC }^y = D_{ CA}^y $.
The i-DMI energy density in the C-plane configuration can then be written as\cite{dmi-mn3sn-2}
\begin{equation}
	u_{\rm idmi} = D \left( \vec e_y \cdot \vec m_A \times \vec m_B 
	                                   + \vec e_{ - \frac{ 5\pi }{ 6 } } \cdot \vec m_B \times \vec m_C
	                                   + \vec e_{ - \frac{ \pi }{ 6 } } \cdot \vec m_C \times \vec m_A
	                           \right) .
\label{idmi_c}
\end{equation}
The magnitude of the i-DMI vectors, $ D $, is common for the $ AB $, $ BC $ and $ CA $ couplings, while their directions lie in the Kagome plane making a regular triangle.
The directions of the i-DMI vectors are shown in Fig.~4.

As a result of the i-DMI in Eq.~(\ref{idmi_c}), the sublattice magnetizations $ \vec m_\zeta $ cant away from the Kagome ($xy$) plane in the ground states, in a way that the canting direction depends on the sublattice so that the $ z $ component of the net magnetization $ \vec m $ remains vanishingly small.
This is consistent with the previous theoretical prediction based on a tight-binding model simulation for Pt/Mn$_3$Sn in the C-plane configuration \cite{dmi-mn3sn-5}.
This prediction is in a clear contrast to that in Ref.~\cite{dmi-mn3sn-2}, where a {\it nonzero net magnetization} is shown to be generated by the DMI of the form of Eq.(\ref{idmi_c}).
This is because they investigated effects of the DMI on the all-in/all-out triangular configuration (corresponding to $ R_+ $), while we are considering the inverse triangular configuration (corresponding to $ R_- $).

To glimpse dynamical effects of the i-DMI in our C-plane configuration, we numerically simulate the switching from $ \varphi = 0 $ to $ \pi $ (two of the six ground states under the intrinsic sixfold anisotropy) under an external field applied in the $ + x $ direction [Fig.~5~(a)].
Note that $ \vec n_1 $ and $ \vec m $ exhibit the opposite $ x $ components in equilibrium.
Shown in Fig.~5~(b) is the numerically-obtained switching field $ \mu_0 h_x^* $ against $ | D | $.
It is seen that, in the C-plane configuration, the i-DMI has negligible impact on the magnetic anisotropy for $ \varphi $.
The switching field of $ \simeq 30 $ mT, little dependent on $ | D | $, predominantly originates from the intrinsic sixfold anisotropy.

So far, we have considered spatially homogeneous systems, i.e., $ \partial_\mu \vec m_\zeta = 0 $ $ ( \mu = x, y, z $ and $ \zeta = A, B, C ) $.
In the presence of the spatial inhomogeneity, $ u_{\rm idmi} $ in Eqs.~(\ref{idmi}) and (\ref{idmi_c}) acquires in general additional terms that depend on $ \partial_\mu \vec m_\zeta $.
For the simplicity of modeling, let us consider the M-plane configuration with the magnetic moments varying their directions only in the $ x $-direction, i.e., $ \partial_z \vec m_\zeta = 0$ $\forall \zeta$.
By requiring the mirror symmetries under the operations $\sigma_z$ and $\sigma_x$ as before, and incorporating the terms at the first order in $\partial_x$, we obtain the inhomogeneous part of the i-DMI energy density, $ u_{\rm idmi}^{\rm inhomo} =  \sum_{ ( \zeta \eta ) } D'_{ \zeta \eta } \vec e_z \cdot \left( \vec m_\zeta \times \partial_x \vec m_\eta + \vec m_\eta \times \partial_x \vec m_\zeta \right) $,
where $ D'_{ \zeta \eta } $ are phenomenological inhomogeneous i-DMI constants.
Assuming that $ D'_{ \zeta \eta } $ have a sufficiently mild dependence on $ ( \zeta \eta ) $, this inhomogeneous part would impart a fixed chirality to the sublattice magnetizations in a domain wall depending on the sign of $ D'_{ \zeta \eta } $, as in essentially the same fashion as in ferromagnets \cite{dmi_dw1, dmi_dw2, dmi_dw3}.

\begin{figure}[t]
\centering
\includegraphics[width=8.5cm, bb=0 0 1070 925]{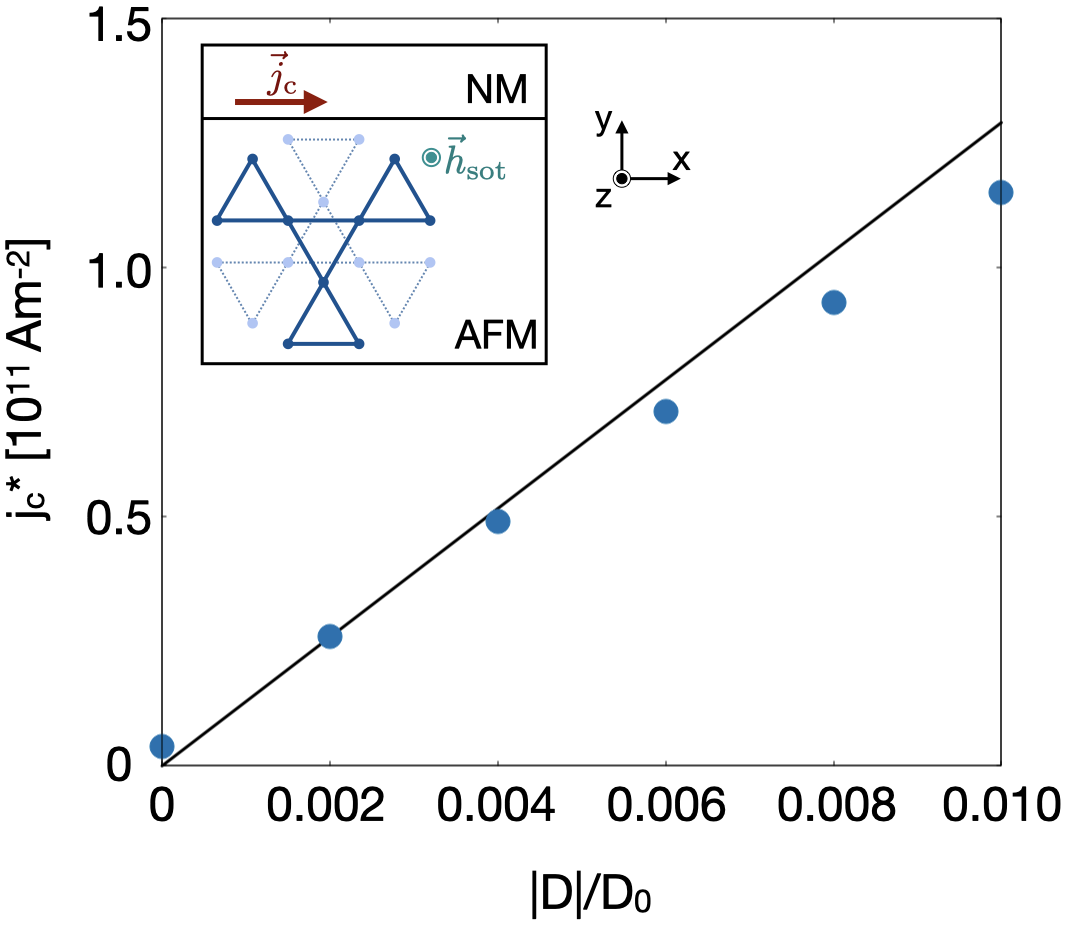}
\caption{
The threshold current density $ j_{\rm c}^* $ for the rotational motion of $ \varphi $, plotted against the absolute value of the i-DMI constant $ | D | $ (normalized by the bulk DMI constant $ D_0 $) in the M-plane configuration.
The solid curve plots Eq.~(\ref{jc}), and the circle symbols are results of the LLG simulations.
See the main text for the parameter values employed.
In the inset, the setup for the current-driven excitation of the AFM is schematically shown.
}
\label{fig06}
\end{figure}

We investigated the field-induced switching in the presence of the i-DMI above.
Here we examine impacts of the i-DMI on the current-driven dynamics of the AFM order.
We consider the spatially-uniform AFM order in the M-plane configuration, with the electric current flowing in the $ x $ direction (see the inset of Fig.~6).
The spin Hall effect in the nonmagnet induces a spin accumulation at the interface, leading to the spin-orbit torque (SOT) exerted on the magnetizations in the adjacent magnet \cite{she}.
The SOT is incorporated in the present model by the torque term
\begin{equation}
	\vec T_{\rm sot}^\zeta = - \gamma \vec m_\zeta \times \left( \vec m_\zeta \times \vec h_{\rm sot} \right) ,
	\qquad ( \zeta = A, B, C ) ,
\label{sot}
\end{equation}
added to the rhs of Eq.~(\ref{llg}), where
\begin{equation}
	\vec h_{\rm sot} = \frac{ 1 }{ 3 } \frac{ \hbar \xi j_{\rm c} }{ 2 e \mu_0 M_S^{\rm sub} d } \vec e_z .
\label{h_sot}
\end{equation}
This is the so-called (anti)damping-like SOT.
In general, there also exists the field-like SOT, which we omit here because it mathematically plays exactly the same role as the external magnetic field in the LLG equations.
Here, $ e $ is the elementary electric charge, $ d $ is the thickness of the AFM thin film, $ \xi $ is the spin Hall angle of the nonmagnet, and $ j_{\rm c} $ is the electric current density.
The factor $ \frac{ 1 }{ 3 } $ in Eq.~(\ref{h_sot}) reflects the assumption that the spin current injected into the AFM transfers its angular momentum equiprobably to each of the three magnetic sublattices.
In this configuration, the SOT can directly couple to $ \varphi $ regardless of the presence/absence of $ \vec m $, and drive a rotational motion of the chiral AFM structure \cite{mn3sn_rot1, mn3sn_rot2, mn3sn_rot3, yamane2019}.
With the i-DMI-induced anisotropy in Eq.~(\ref{u_tot}), the equation of motion for $ \varphi $ is given by $ ( 3 \gamma J / \mu_0 M_S^{\rm sub} )^{-1} \partial_t^2 \varphi + \alpha \partial_t \varphi = - ( \gamma K_{\rm idmi} / 3 \mu_0 M_S^{\rm sub} )\sin2\varphi + \gamma h_{\rm sot}^z $.
By equating the two ``forces'' in the rhs originating from the i-DMI and the SOT, the threshold current density $ j_{\rm c}^* $ for the rotational motion of $ \varphi $ can be defined as
\begin{equation}
	| j_{\rm c}^* |
	= \frac{ 2 e d }{  \hbar | \xi | } K_{\rm idmi}
	= \frac{ 4 e d }{ \sqrt{ 3 } \hbar | \xi | } \frac{  K }{  J } | D | .
\label{jc}
\end{equation}
Shown in Fig.~6 is $ j_{\rm c}^* $ as a function of $ | D | $, where the analytical and numerical results are compared.
We employed $ d = 10 $ nm, $ \xi = 0.1 $, and the same values for the other parameters as in the simulation in Fig.~3.
At $ D = 0 $, the small but finite threshold current density $ j_{\rm c}^* \simeq 4 \times 10^{9} $ Am$^{-2}$ is due to the intrinsic sixfold anisotropy.

The results shown in Fig.~6 serve as yet another confirmation that the i-DMI in the M-plane configuration provides a uniaxial anisotropy for $ \varphi $, being consistent with the analytical prediction Eq.~(\ref{u_idmi}).
We note that, by Eqs.~(\ref{sot}) and (\ref{h_sot}), we have assumed that the SOT can be solely attributed to the spin Hall effect in the nonmagnet, and $ \vec h_{\rm sot} $ is common to all the three magnetic sublattices.
In general, the Rashba spin-orbit coupling at the interface can also contribute to the SOT \cite{rashba1, rashba2}.
Furthermore, the broken inversion and sublattice symmetries at the interface allow for the SOT to be asymmetric among the three magnetic sublattices.
It is beyond the scope of this work to address these issues, which will be discussed elsewhere.

\section{Conclusion}
In conclusion, we have theoretically formulated the DMI appearing at the interface of a nonmagnetic/noncollinear-AFM thin film heterostructure.
By exploiting symmetry arguments at the interface, we have derived analytical expressions for the i-DMI in a continuum, three magnetic-sublattice model.
We conclude that the i-DMI causes a uniaxial magnetic anisotropy for the AFM order parameter in the M-plane configuration, even in the absence of interfacial anisotropy for each magnetic moment at the microscopic level.
Such an effect of the i-DMI may be partially responsible for the uniaxial anisotropy experimentally reported recently in Mn$_3$Sn thin-film systems \cite{mn3sn_tp4, mn3sn_switch3, mn3sn_switch7, mn3sn-ani1, mn3sn_epi3}, a quantitative confirmation of which would require controlled experiments combined with microscopic theoretical approaches.
Our finding also suggests the possibility of manipulating the magnetic anisotropy in noncollinear AFMs via the i-DMI.

{\it Acknowledgments ---}
The authors are grateful for valuable comments from T. Uchimura, T. Dohi, Y. Takeuchi, H. Ohno, and J. Ieda.
This work was supported by JSPS KAKENHI (No. 22K03538, 22KK0072, 23H01828, 24H00039 and 24H02235), JST TI-FRIS, MEXT X-NICS (No. JPJ011438), and RIEC Cooperative Research Projects.


\end{document}